\documentclass[prl,twocolumn,showpacs,amsfonts,amsmath,floatfix]{revtex4}
\usepackage{graphicx}
\usepackage{psfrag}
\newcommand{\Journal}[4]{#1 {\bf #2}, #3 (#4)}
\newcommand{\PR}{Phys. Rev.}
\newcommand{\PRL}{Phys. Rev. Lett.}
\newcommand{\PRA}{Phys. Rev. A}

\newcommand{\JMP}{J. Math. Phys.}

\newcommand{\Science}{Science}
\newcommand{\PLA}{Phys. Lett. A}
\begin{document}
\title {Pairing, off-diagonal long-range order, and quantum phase transition 
in strongly attracting ultracold Bose gas mixtures in tight waveguides}
\author{M. D. Girardeau}
\email{girardeau@optics.arizona.edu}
\affiliation{College of Optical Sciences, University of Arizona,
Tucson, AZ 85721, USA}
\date{\today}
\begin{abstract}
A model of two 1D ideal Bose gases A and B with strong AB attractions induced
by a p-wave AB Feshbach is studied. The model is solved exactly by a
Bose-Bose duality mapping, and it is shown that  
there is no A-component or B-component Bose-Einstein condensation
and no AB-pair off-diagonal long-range order (ODLRO),
but both AA-pair and BB-pair ODLRO. After generalization by adding
even-wave AA and BB repulsions and reducing the strength of the odd-wave
AB attraction by Feshbach resonance detuning, a quantum phase transition 
occurs between a phase with AB contact nodes and one with no such nodes. 
\end{abstract}
\maketitle

Strong interatomic interactions and correlations occur in ultracold gases
confined in de Broglie waveguides with transverse trapping
so tight that the atomic dynamics is essentially one-dimensional (1D)
\cite{Ols98}, with confinement-induced resonances \cite{Ols98,GraBlu04}
allowing Feshbach resonance tuning \cite{Rob01} of the effective 1D
interactions to very large values.  This has led to experimental verification
\cite{Par04Kin04,Kin05,Kin06} of the fermionization of bosonic ultracold
vapors in such geometries predicted by the Fermi-Bose (FB)
mapping method \cite{Gir60Gir65}, an exact mapping of a 1D gas of bosons with
point hard core repulsions, the ``Tonks-Girardeau'' (TG) gas, to an
\emph{ideal} spin-aligned Fermi gas. The ``fermionic Tonks-Girardeau'' (FTG)
gas \cite{GirOls04,GirNguOls04}, a 1D
spin-aligned Fermi gas with very strong \emph{attractive}
interactions, can be realized by a 3D p-wave Feshbach resonance as, e.g., in
ultracold $^{40}$K vapor \cite{Tik04}. It has been pointed out
\cite{CheShi98,GraBlu04,GirOls04,GirNguOls04} that the FB mapping
\cite{Gir60Gir65} can be exploited to map the
FTG gas to the \emph{ideal} Bose gas. The very strong fermion-fermion 
attraction in the FTG gas leads to fermion pairing and superconductive
off-diagonal long-range order (ODLRO) of the two-fermion  
density matrix \cite{GirMin06}.

There are a number of models of strongly interacting 1D ultracold gas mixtures 
which are exactly soluble by generalizations of the FB mapping \cite{GirMin07}.
In \cite{GirMin07,FanVigMinMin08} the properties of one such soluble model,
a mixture of a TG Bose gas A and an ideal Fermi gas B, with point hard-core AB 
interactions, were investigated in detail. Here I shall discuss another
model, a mixture of two different ideal Bose gases A and B, with an
AB interaction of FTG form. This model is exactly soluble, and 
it will be shown that it has very unusual behavior: The strong AB attraction 
destroys the ground state Bose-Einstein condensation (BEC) 
and single-particle off-diagonal long-range order (ODLRO) 
of both components A and B, and it induces both AA and BB pairing
manifested in superconductive ODLRO of both the two-A and two-B density
matrices, although there are no AA or BB interactions. Furthermore, 
there is no AB pair ODLRO in spite of the strong AB attractions. 
It will also be shown that if the AB attraction is a finite odd-wave 
attraction rather than
the infinite FTG limit, and there is also a repulsive even-wave AB interaction
of Lieb-Liniger (LL) delta function form 
\cite{LieLin63}, then there is a quantum phase transition as the coupling
constants are varied, between a phase in which there are no AB contact nodes
and only the repulsive LL interaction acts, and another phase in which
there are AB contact nodes and only the attractive FTG-like interaction acts.

{\it FTG interaction and nodal structure:} The FTG gas is a spin-aligned 1D
Fermi gas with infinitely strongly attractive zero-range odd-wave interaction 
induced by a p-wave Feshbach resonance. It is the infinite 1D scattering
length limit $a_{1D}\to -\infty$ of a 1D Fermi gas with zero-range
attractive interactions leading to a 1D scattering length defined 
in terms of the ratio of the derivative $\Psi^{'}$ of the wave function
to its value at contact: 
$\Psi(x_{jk}=0+)=-\Psi(x_{jk}=0-)= -a_{1D}\Psi^{'}(x_{jk}=0\pm)$ where the 
prime denotes the derivative with respect to $x_{jk}$ 
\cite{GraBlu04,GirOls04,GirNguOls04}.
The FTG limit $a_{1D}\to -\infty$ corresponds to a 1D zero-energy odd-wave
scattering resonance reachable by Feshbach resonance tuning to a 1D odd-wave
confinement-induced resonance \cite{Rob01,Ols98,GraBlu04}. There are
several different zero-range pseudopotentials which generate this contact
condition. One representation \cite{GirOls04} is 
$\hat{v}_o=g_o\delta^{'}(x_{jk})\hat{\partial}_{\pm}$ where
$\hat{\partial}_{\pm}\Psi(x_{jk})=(1/2)[\Psi^{'}(0+)+\Psi^{'}(0-)]$.
The FTG limit $a_{1D}\to -\infty$ is equivalent to $g_o\to +\infty$.
This representation explicitly exhibits
an odd-wave projection property of the interaction, i.e., it vanishes on even 
functions of $x_{jk}$. 

Since there is no particular symmetry
under exchange of particles of the different species A and B, AB scattering
in all partial waves is possible, but usually s-wave scattering dominates.
However, in the neighborhood of a p-wave AB resonance, p-wave scattering
dominates, and gives rise in 1D to an odd-wave effective interaction of
FTG form \cite{GraBlu04,GirOls04,GirNguOls04}. For this it is important to 
realize that the odd-wave projection property of $\hat{v}_o$ requires only 
\emph{local} antisymmetry,
i.e., it guarantees that if $x_i$ is an A-particle 
position and $y_j$ a B-particle position, then $\hat{v}_o\Psi(x_i,y_j)$ is 
nonzero only if
$\Psi$ has a node at $x_i=y_j$, where $\Psi(x_i,y_j)=-\Psi(y_j,x_i)$
in an \emph{infinitesimal} neighborhood of the node. It is not
necessary that there be \emph{global} antisymmetry under such exchange,
and indeed, the ground state $\Psi_0$ derived in the following sections has no 
such global antisymmetry. In the presence of both a 1D even-wave 
AB Lieb-Liniger
interaction $g_e\delta(x_i-y_j)$ \cite{LieLin63} generated by 3D s-wave
AB scattering and a 1D odd-wave resonance generated
by a 3D p-wave Feshbach resonance for AB scattering, the wave function
can lower its energy by developing nodes at $x_i=y_j$ so as to 
kill the repulsive even-wave interaction and activate the 
strong 1D odd-wave attraction. 

The contact discontinuities of $\Psi$ \cite{CheShi98} can also be understood 
as a zero-range limit $x_0\to 0+$ and $V_0\to\infty$ of the two-body scattering
solution for a square well of width $2x_0$ and depth $V_0$, where the limit is 
carried out such that $V_0 x_0^2$ approaches a finite, nonzero limit 
\cite{GirOls04,GirNguOls04}. In the untrapped  case, the exterior solution
for scattering length $a_{AB}\to -\infty$ is constant (+1 for $x-y>0$ and -1 
for $x-y<0$), and the interior solution is $\sin[\kappa(x-y)]$ with 
$\kappa=\sqrt{2\mu V_{0}/\hbar^2}=\pi/2x_0$ and $\mu$ the effective mass
$m_Am_B(m_A+m_B)$; hence $\kappa x_0=\pi/2$ and 
$V_0x_0^2=(\pi\hbar)^2/8\mu$. 
In the zero-range limit the interior kinetic energy $\to +\infty$ and
potential energy $\to -\infty$, but their sum remains zero, the ground state
energy. Since the FTG interaction acts only on odd waves, the FTG interaction
$\hat{v}_0$ should be written as $\hat{v}_0=v\hat{P}_0$ where $v$ is the above 
square well and $\hat{P}_0$ is the odd-wave projector.

{\it Many-body ground state:} Assuming trapping on a ring, 
the Hamiltonian consists only of the kinetic energy
operators of components A and B plus the AB FTG interaction:
\begin{equation}\label{H}
\hat{H}=\sum_{i=1}^{N_A}\frac{-\hbar^2}{2m_A}
\frac{\partial^2}{\partial x_i^2}
+\sum_{i=1}^{N_B}\frac{-\hbar^2}{2m_B}\frac{\partial^2}{\partial y_i^2}
+\sum_{i=1}^{N_A}\sum_{j=1}^{N_B}\hat{v}_o(x_i-y_j)
\end{equation}
where $\hat{v}_o$ is the previously defined odd-wave FTG interaction. 
The wave functions satisfy periodic boundary conditions with periodicity length
$L$ (the ring circumference) with respect to all the $x_i$ and $y_i$. The
scattering length $a_{AB}$ is $-\infty$ in the FTG limit. For square well 
width $2x_0$ nonzero but very small, the unnormalized ground state $\Psi_0$ is 
constant (say $\pm 1$) when all $|x_i-y_j|>2x_0$, except for sign changes as 
each $x_i-y_j$ varies from $-x_0$ to $x_0$ in accordance with the internal 
wave function $\pm\sin[\kappa(x_i-y_j)]$. The condition 
$\kappa=\sqrt{2\mu V_0/\hbar^2}=\pi/2x_0$ determines the well depth $V_0$
such that the scattering length $a_{AB}$ is $-\infty$, and the FTG limit is 
obtained by letting $x_0\to 0$ and $V_0\to\infty$ in accordance with this
condition. In this limit the internal
wave function becomes invisible and $\Psi_0$ appears to jump discontinuously
between $\pm 1$ whenever an A-particle passes a B-particle, but there are
hidden nodes at $x_i-y_j=0$ at the centers of the wells. The ground state 
energy $E_0=0$, generalizing the situation for the pure FTG 
gas \cite{GraBlu04,GirOls04,GirNguOls04,GirMin06}. $\Psi_0$ maps to a 
``model state'' $\Psi_{M0}$ consisting of two noninteracting ideal Bose gases 
totally Bose-Einstein condensed into their ground orbital, which is a trivial
constant for periodic boundary conditions: $\Psi_0=\Psi_{M0}M=M$ and
$\Psi_{M0}=1$ where $M$ is the mapping function
\begin{equation}\label{M}
M(x_1,\cdots,x_{N_A};y_1,\cdots,y_{N_B})
=\prod_{i=1}^{N_A}\prod_{j=1}^{N_B}\text{sgn}(x_i-y_j)
\end{equation}
where the sign function $\text{sgn}(x)$ is $+1\ (-1)$ if $x>0\ (x<0)$. 
Although the mapped bosonic state $\Psi_{M0}$ is a trivial 
constant outside the square wells, the interior wave function  
$\pm\sin(\kappa|x_i-y_j|)$ vanishes with cusps at $x_i-y_j=0$.
Therefore, physical consistency requires the presence of a zero-diameter
hard core interaction added to the square well. The mapped Bose
gas is then not truly ideal, but rather a \emph{TG gas with 
superimposed attractive well}, whose nontrivial interior wave function becomes 
invisible in the zero-range limit, simulating a mixture of two noninteracting 
ideal Bose gases insofar as the energy and exterior wave function are 
concerned. The densities of components A and B are trivial constants in the 
ground state,
but the off-diagonal elements of the reduced density matrices of $\Psi_0$
are quite nontrivial and interesting, due to the effects of the discontinuities
in $M$; they will be discussed in later sections. 

The periodic boundary 
conditions impose constraints on the values of $N_A$ and $N_B$. 
Suppose that the positions of all B-bosons and all but one A-boson,
say $x_1$, are fixed on the open interval (0,L). Starting with that A-boson 
at $x_1>0$ but to the left of all the other particles and moving it to 
a position $x_1<L$ but to the right of all the others, one counts $N_B$
sign changes of $\Psi_0$ due to the AB contact nodes, so $N_B$ 
must be even for $L$-periodicity in the A-boson coordinates.
Repeating this process with all A-bosons and 
all but one B-boson fixed and moving that B-boson instead, one counts $N_A$
sign changes; hence $N_A$ must be even for $L$-periodicity in the
B-boson coordinates. 

{\it One-particle density matrices and momentum distributions:} Generalizing 
the derivation in \cite{GirWri05,BenErkGra04,GirMin06}, one finds that the 
one-particle density matrix of component A is 
\begin{widetext}
\begin{equation}\label{rho1A}
\rho_{1A}(x,x')=N_AL^{-N_A-N_B}
\int\Psi_0(x,x_2,\cdots,x_{N_A};Y)
\Psi_0(x',x_2,\cdots,x_{N_A};Y)dx_2\cdots dx_{N_A}
dY=n_A[I(x,x')]^{N_B}
\end{equation}
\end{widetext}
where $Y=(y_1,\cdots, y_{N_B})$, $n_A=N_A/L$, and 
$I(x,x')=L^{-1}\int_{-L/2}^{L/2}\text{sgn}(x-\xi)\text{sgn}(x'-\xi)d\xi
=1-2|x-x'|/L$. In the thermodynamic limit where $N_B\to\infty$ and
$L\to\infty$ such that $N_B/L\to n_B$ with finite and nonzero
number density $n_F$ $n_B$, one has
$[I(x,x')]^{N_B}\to e^{-2n_B|x-x'|}$ in analogy with the one-component
FTG gas case \cite{GirWri05,BenErkGra04,GirMin06}. Hence 
$\rho_{1A}(x,x')=n_Ae^{-2n_B|x-x'|}$, and by interchange of A and B
$\rho_{1B}(y,y')=n_Be^{-2n_A|x-x'|}$. Their Fourier transforms $n_{kA}$ and
$n_{kB}$, normalized to $\sum_{k}n_{kA}=N_A$ and $\sum_{k}n_{kB}=N_B$ where
the allowed momenta are $k=\nu 2\pi/L$ with $\nu=0,\pm 1,\pm 2,\cdots$),
are Lorentzian discrete momentum distributions
$n_{kA}=\frac{4n_An_B}{4n_B^2+k^2}$ and $n_{kB}=\frac{4n_An_B}{4n_A^2+k^2}$.
The Fermi-like Lorentzian shapes are strong modifications of the ideal Bose 
gas distributions $N_A\delta{k0}$ and $N_B\delta{k0}$ due to the infinite AB 
attraction; no trace of BEC of components A and B, and of the associated ODLRO 
of $\rho_{1A}$ and $\rho_{1B}$, remains. Nevertheless, at $k=0$
$n_{kA}$ reduces to $n_A/n_B$ which increases without limit as the B-component
density $n_B$ falls to zero for fixed $n_A$, and in fact the continuous 
momentum distribution $(L/2\pi)n_{kA}$ reduces to a representation of the ideal
Bose gas distribution $N_A\delta(k)$ as $n_B\to 0$. $n_{kB}$ has the same
behavior, with A and B interchanged.

{\it Two-particle density matrices, pairing, and ODLRO:} The two-particle 
A-component and B-component density matrices $\rho_{2AA}(x_1,x_2;x_1',x_2')$ 
and $\rho_{2BB}(y_1,y_2;y_1',y_2')$ can also be evaluated in closed form
by generalization of (\ref{rho1A}) and the derivation for the one-component 
FTG gas in \cite{GirMin06}. For component A 
\begin{eqnarray}
&&\rho_{2AA}(x_{1},x_{2};x_{1}',x_{2}')=N_A(N_A-1)L^{-(N_A+N_B)}\nonumber\\
&\times&\int\Psi_0(x_1,x_2,x_3,\cdots,x_{N_A};Y)\nonumber\\
&\times&\Psi_0(x_1',x_2',x_3,\cdots,x_{N_A};Y)dx_3\cdots dx_{N_A}dY
\end{eqnarray}
Using 
$\Psi_0(x_1,x_2,x_3,\cdots,x_{N_A};Y)=\prod_{j=1}^{N_B}
\text{sgn}(x_1-y_j)\text{sgn}(x_2-y_j)$ one finds in the thermodynamic limit
$\rho_{2AA}(x_{1},x_{2};x_{1}',x_{2}')=n_A^2e^{2n_B(z_1-z_2+z_3-z_4)}$
where $z_{1}\le z_{2}\le z_{3}\le z_{4}$ are the arguments 
$(x_{1},x_{2};x_{1}',x_{2}')$ in ascending order. If $x_1<x_2<x_1'<x_2'$ then
$\rho_{2AA}=n_A^2e^{-2n_B|x_1-x_2|}e^{-2n_B|x_1'-x_2'|}$.  
Generalizing the argument in \cite{GirMin06} one sees that if the variable 
pairs $(x_1,x_2)$ and $(x_1',x_2')$ are separated to arbitrary distance
while keeping $|x_1-x_2|$ and $|x_1'-x_2'|$ fixed, then $\rho_{2AA}$
remains constant, signalling AA-pair ODLRO associated with a 
leading term $\lambda_{1AA}u_{1A}(x_1,x_2)u_{1A}(x_1',x_2')$ in the spectral
representation of $\rho_{2AA}$, with eigenfunction
$u_{1A}(x_1,x_2)=\mathcal{C}_A\ e^{-2n_B|x_1-x_2|}$, normalization
constant $\mathcal{C}_A=\sqrt{2n_B/L}$, and macroscopic eigenvalue
$\lambda_{1AA}=n_A^2/\mathcal{C}_A^2=n_AN_A/2n_B$. There is a BEC-BCS
crossover from AA-pair BEC when $n_B\gg n_A$ and the range of $u_{1A}$
is $\ll 1/n_A$ implying tightly bound AA pairs, to AA-pair superconductivity
when $n_B\ll n_A$ and the range of $u_{1A}$ is $\gg 1/n_A$ implying
extended and strongly overlapping AA Cooper pairs. Since 
$\rho_{2BB}(y_{1},y_{2};y_{1}',y_{2}')$ exhibits the same behavior with
A and B interchanged, one concludes that when $n_B\gg n_A$ there is coexistence
of BEC of AA pairs and superconductivity of BB pairs, and when $n_B\ll n_A$
The AA and BB pairing is a purely off-diagonal phenomenon, both in the case
of superconductive ODLRO with weakly bound Cooper pairs and in the case
of BEC of tightly-bound pairs. There is no diagonal AA, BB, or AB order;
the pair distribution functions 
$D_{AA}(x_1,x_2)=n_A^{-2}\rho_{2AA}(x_{1},x_{2};x_{1},x_{2})$,
$D_{BB}(y_1,y_2)=n_B^{-2}\rho_{2BB}(y_{1},y_{2};y_{1},y_{2})$,
and $D_{AB}(x,y)=(n_An_B)^{-1}\rho_{2AB}(x,y;x,y)$ are all constant,  
as is most easily seen
by noting that (a) our system of A-bosons and B-bosons with FTG AB attraction
maps to a mixture of ideal A-Bose and B-Bose gases with no AB interaction,
and (b) \emph{diagonal} density matrix elements are invariant under 
mapping via Eq. (\ref{M}). This generalizes the previous result for the pure 
FTG gas, where there is superconductive ODLRO but the pair distribution 
function is constant \cite{GirMin06}. 

The two-particle AB density matrix is also of interest. By a derivation
paralleling that for $\rho_{2AA}$ and $\rho_{2BB}$ one finds in the 
thermodynamic limit
\begin{eqnarray}\label{rho_AB}
\rho_{2AB}(x,y;x',y')&=&n_An_B\text{sgn}(x-y)\text{sgn}(x'-y')\nonumber\\
&\times&e^{-2n_B|x-x'|}e^{-2n_A|y-y'|}\ .
\end{eqnarray}
Suppose that $x'=x+d$ and $y'=y+d$. Then 
$\rho_{2AB}=n_An_Be^{-2(n_A+n_B)d}$ which vanishes exponentially as
$d\to\infty$. It follows that there is no AB-pair ODLRO in spite of the
strong AB attraction. It is informative in this connection to compare and 
contrast two cases (a) two ideal Bose gases A and B with no AB 
interactions, and (b) the present case, two ideal Bose gases A and B with FTG 
AB interactions. In case (a) there is complete BEC of both components A and B,
the many-body ground state is a trivial constant, $\rho_{2AA}$, $\rho_{2BB}$,
and $\rho_{2AB}$ are also constant, and hence all three of these density
matrices exhibit ODLRO. However, the AA and BB ODLRO is a trivial consequence
of the trivial ground state structure and more generally, follows from the
ODLRO of $\rho_{1A}$ and $\rho_{1B}$ \cite{Yan62} and implies no true AA or BB
pairing correlation. In the present case (b), there is no BEC of component
A or B and hence no ODLRO of $\rho_{1A}$ or $\rho_{1B}$, but there is ODLRO
of both $\rho_{2AA}$ and $\rho_{2BB}$, implying both AA and BB pairing, 
since $u_{1A}$ and $u_{1B}$ have finite range. On the other hand, $\rho_{2AB}$
has no ODLRO.

{\it Quantum phase transition:} Suppose now that in addition to the odd-wave
AB interaction of FTG form, there is also an even-wave AB 
interaction of LL delta function form \cite{LieLin63},
$v_e(x_i-y_j)=g_{AB}\delta(x_i-y_j)$ with $g_{AB}>0$ (repulsive interaction).
Recall that before passing to the infinitely narrow well limit
$x_0\to 0$, $V_0\to\infty$ of the FTG interaction, the ground state in the case
of no even-wave interaction has nodes at AB contact due to the internal
wave function $\sin\kappa(x_i-y_j)$. These nodes kill the even-wave 
interaction, so that the ground state in the presence of the even-wave
interaction is the same as that in the absence of an even-wave interaction,
which has energy zero. Any state without AB contact nodes has positive
energy in the presence of the even-wave repulsion, since then the odd-wave
FTG interaction is killed; the system develops AB contact nodes
spontaneously in the presence of the FTG odd-wave attraction in order to lower 
its energy by killing the even-wave repulsion. Now suppose that there is no
even-wave interaction, but the odd-wave attraction is weakened by carrying out 
the limit $x_0\to 0$ and $V_0\to\infty$ in such a way as to
produce a large but finite negative odd-wave scattering length,
$-\infty<a_{ABo}<0$. So far the FTG limit where $\kappa x_0=\frac{\pi}{2}$ and 
$a_{ABo}=-\infty$ has been assumed, but more generally if $a_{ABo}$ is negative
and finite, then $\kappa x_0=\frac{\pi}{2}-\frac{2x_0}{\pi|a_{ABo}|}$ as
$x_0\to 0$ \cite{GirOls03,GirNguOls04}. The ground state $\Psi_0$ in that
case maps via (\ref{M}) to a model ground state $\Psi_{M0}$ 
consisting of two Bose gases A and B with no AA or BB interaction but an
even-wave AB interaction of LL form $g_{AB}'\delta(x_i-y_j)$ with
$g_{AB}'=\hbar^2/\mu|a_{ABo}|$ and positive energy. If one generalizes
further by adding a nonzero even-wave interaction $g_{AB}\delta(x_i-y_j)$
again, then so long as the ground state $\Psi_0$ is retained unchanged
with AB contact nodes, the even-wave interaction will be killed and will
have no effect. However, if $g_{AB}<g_{AB}'=\hbar^2/\mu|a_{ABo}|$,
then a ground state with no AB contact nodes has lower energy, since
then it is the odd-wave interaction which is killed, instead of the
even-wave interaction. It follows that there is a quantum phase transition
between a phase with no AB contact nodes, which has lower energy when
$g_{AB}<g_{AB}'$, and a phase with AB contact nodes, which has lower
energy when $g_{AB}>g_{AB}'$. This is very similar to the 
ferromagnetic-antiferromagnetic phase transition in the 1D
spinor Fermi gas with both even and odd-wave interactions \cite{Note1,Gir06}.

{\it Stronger AB attraction:} It was pointed out above that if 
$-\infty<a_{ABo}<0$ then $\kappa x_0=\frac{\pi}{2}-\frac{2x_0}{\pi|a_{ABo}|}$ 
as $x_0\to 0$, or without the absolute value signs
$\kappa x_0=\frac{\pi}{2}+\frac{2x_0}{\pi a_{ABo}}$. In analogy with
the case of fermions discussed recently \cite{GirWri08}, for an interaction of
FTG form the same relation (without absolute value signs) holds if 
$a_{ABo}>0$, in which case the model state $\Psi_{M0}$ generated by the mapping
(\ref{M}) has an \emph{attractive} LL AB interaction
$g_{AB}\delta(x_i-y_j)$ with
$g_{AB}=-\hbar^2/\mu a_{ABo}<0$ and \emph{negative} energy. 
The $N=2$ model ground state is bound, $\Psi_{0M}=e^{-|x-y|/a_{ABo}}$ 
with energy $E_0=-\frac{\hbar^2}{2\mu a_{ABo}^2}$, and the corresponding actual
physical state is $\Psi_0=\text{sgn}(x-y)e^{-|x-y|/a_{ABo}}$, with the same
energy. The exact solution for both $N_A>2$ and $N_B>2$ is not known, but 
if $\Psi_{M0}$ has Bijl-Jastrow form
$\Psi_{M0}=\mathfrak{N}\prod_{i=1}^{N_A}\prod_{j=1}^{N_B}f(|x_i-y_j|)$ where
$\mathfrak{N}$ is a normalization constant and 
$f(|x_i-y_j|)\approx e^{-|x_i-y_j|/a_{ABo}}$ for 
$|x_i-y_j|\ll(\sqrt{n_An_B})^{-1}$, then in the neighborhood of each 
$x_i-y_j=0$ the wave function reduces to that of an AB dimer. The behavior
of $\Psi_{M0}$ as $x_i$ recedes to distances $x_i\gg(\sqrt{n_An_B})^{-1}$ from
$y_j$ will be controlled by proximity of $x_i$ to other B-particles, not
$y_j$. If $f(\xi)\to\lambda$ 
for $\xi\gg(\sqrt{n_An_B})^{-1}$ where $\lambda$ is some nonzero constant,
then $\mathfrak{N}=\lambda^{-N_B}$. The physical state $\Psi_0$
generated by the mapping (\ref{M}) then has reduced density matrices
reducible to 1D integrals as previously. However, in the thermodynamic
limit $\rho_{2AB}$ reduces to the previous expression (\ref{rho_AB}),
with no AB-pair ODLRO. A better approximation to $\Psi_0$
might reverse this conclusion, so this model deserves further study.

{\it Outlook:} In view of the surprising properties of this model, 
experimental study of mixtures of weakly interacting A-bosons
and B-bosons in tight waveguides with strong 1D AB attractions \cite{GraBlu04} 
induced by a p-wave AB Feshbach resonance \cite{Tik04} should be 
fruitful. In particular, one could look for the predicted quantum phase 
transition.    
\begin{acknowledgments}
I thank Peter Reynolds and Gregory Astrakharchik for helpful comments on drafts
of this work.
\end{acknowledgments}

\begin{thebibliography}{19}
%
\bibitem{Ols98} M. Olshanii, \Journal{\PRL}{81}{938}{1998}.
%
\bibitem{GraBlu04} B.E. Granger and D. Blume, \Journal{\PRL}{92}{133202}{2004}.
%
\bibitem{Rob01} J.L. Roberts {\it et al.}, \Journal{\PRL}{86}{4211}{2001}.
%
\bibitem{Par04Kin04} B. Paredes, {\it et al.},
Nature {\bf 429}, 277 (2004); T. Kinoshita, T.R. Wenger, and D.S. Weiss,
\Journal{\Science}{305}{1125}{2004}.
%
\bibitem{Kin05} T. Kinoshita, T.R. Wenger, and D.S. Weiss,
\Journal{\PRL}{95}{190406}{2005}.
%
\bibitem{Kin06} T. Kinoshita, T.R. Wenger, and D.S. Weiss, Nature {\bf 440},
900 (2006).
%
\bibitem{Gir60Gir65} M. Girardeau, \Journal{\JMP}{1}{516}{1960};
M.D. Girardeau, \Journal{\PR}{139}{B500}{1965}, Secs. 2, 3, and 6.
%
\bibitem{GirOls04} M.D. Girardeau and M. Olshanii, 
\Journal{\PRA}{70}{023608}{2004}.
%
\bibitem{GirNguOls04} M.D. Girardeau, Hieu Nguyen, and M. Olshanii,
Optics Communications {\bf 243}, 3 (2004).
%
\bibitem{Tik04} C. Ticknor, C.A. Regal, D.S. Jin, and J.L. Bohn,
\Journal{\PRA}{69}{042712}{2004}.
%
\bibitem{CheShi98} T. Cheon and T. Shigehara, \Journal{\PLA}{243}{111}{1998}
and \Journal{\PRL}{82}{2536}{1999}.
%
\bibitem{GirMin06} M.D. Girardeau and A. Minguzzi, 
\Journal{\PRL}{96}{080404}{2006}.
%
\bibitem{GirMin07} M.D. Girardeau and A. Minguzzi, 
\Journal{\PRL}{99}{230402}{2007}.
%
\bibitem{FanVigMinMin08} B.Y. Fang, P. Vignolo,C. Miniatura, and A. Minguzzi,
arXiv:0809.4419.
%
\bibitem{LieLin63} E.H. Lieb and W. Liniger, Phys. Rev. {\bf 130},
1605 (1963).
%
\bibitem{GirWri05} M.D. Girardeau and E.M. Wright, 
\Journal{\PRL}{95}{010406}{2005}.
%
\bibitem{BenErkGra04} S.A. Bender, K.D. Erker, and B.E. Granger, 
\Journal{\PRL}{95}{230404}{2005}.
%
\bibitem{GirOls03} M.D. Girardeau and M. Olshanii, arXiv:cond-mat/0309396.
%
\bibitem{Yan62} C.N. Yang, Rev. Mod. Phys. {\bf 34}, 694 (1962).
%
\bibitem{Note1} See pp. 19-20 of \cite{GirNguOls04}.
%
\bibitem{Gir06} M.D. Girardeau, \Journal{\PRL}{97}{210401}{2006}.
%
\bibitem{GirWri08} M.D. Girardeau and E.M. Wright, 
\Journal{\PRA}{77}{043612}{2008}.
%
\end{thebibliography}
\end{document}